\documentclass[12pt,letterpaper]{article}
\usepackage{amsmath,amssymb,pgf,pgfarrows,pgfnodes,float,appendix, hyperref}
\usepackage{graphicx}
\usepackage{subfigure}
\usepackage[margin=0.9in]{geometry}

\title{{\rm\footnotesize \qquad \qquad \qquad \qquad \qquad \ \qquad \qquad \qquad \ \ \ \ \ \                  RUNHETC-2018-27, UTTG-17-18}\vskip.5in The Gravitational Memory of a Galaxy}
\author{Tom Banks\\
Department of Physics and NHETC\\
Rutgers University, Piscataway, NJ 08854\\
E-mail: \href{mailto:tibanks@ucsc.edu}{tibanks@ucsc.edu}\\ 
W. Fischler\\
Department of Physics \\
University of Texas, Austin,TX, 78712\\
E-mail: \href{mailto:fischler@utexas.edu}{fischler@utexas.edu}}

\date{September 30, 2018}
\begin{document}
\maketitle

\begin{abstract}  We argue that soft gravitational radiation leads to a misidentification of the angular momentum of stars seen in distant galaxies, and that this could be interpreted as an additional mass inside the orbit of the star.  It is tempting to identify this with the modifications of Newtonian dynamics that have been claimed\cite{mond} to eliminate the necessity for dark matter. Simple estimates of the effect show that it is very small and does not have the right functional form to explain the Tully-Fisher relation.  The effect does grow with the age of the universe, but if our universe is indeed dominated by a cosmological constant, then most galaxies will disappear beyond the cosmological horizon before the effect is observable. We conjecture that this effect can be thought of as a new kind of gravitational memory: a rotation of frames at infinity, relative to those near the galaxy. One of the conclusions of our work is that at very large times the linearized classical treatment of this problem is inadequate, and should be replaced by Weinberg's soft photon resummation. However, these times are so long that they do not appear relevant to the universe in which we live. Neither method of calculation leads to a significant effect on galaxy evolution.\end{abstract}

\section{Introduction}

There is a large literature devoted to following up on Milgrom's proposal\cite{MOND1} that the rotation curves of galaxies can be explained by a modification of the laws of gravity, rather than dark matter.  In particular, E. Verlinde has pioneered an approach to such an explanation based on holographic ideas superficially very similar to our own\cite{hst}.  Indeed\cite{verlinde1} Verlinde introduces his equations for the elasticity of space by invoking the decrease of entropy in a causal diamond in the presence of matter.  In Holographic Space-time models, this fundamental observation is the basis for the HST description of localized objects, the resolution of the firewall paradox\cite{fw3} and the HST theory of inflationary and post inflationary cosmology\cite{holocosmrevised}.

Upon closer examination we have found that the HST implementation of this idea does not give rise to Verlinde's elastic equations.  In HST there is indeed a modification of Newton's law, but it is a subleading effect in inverse powers of the impact parameter in a two body collision, such as one might expect from higher derivative terms in the low energy gravitational action.  Furthermore, in the context of a Newtonian orbit, the HST effect is analytic in the central mass.  Verlinde's modifications of Newtonian laws increase with distance, and depend on the square root of the central mass.  In addition, it has become clear that relativistic versions of MOND or of Verlinde's equation require additional low energy fields\cite{relmond}, whereas, following the arguments of Jacobson\cite{ted} HST was constructed to give precisely Einstein's equations (or rather the supersymmetric generalizations of them) in the long wavelength limit.

We were intrigued by the appearance of non-analytic dependence on the central mass, which appears in all versions of MOND in order to implement the Tulley-Fisher relation as a simple consequence of the gravitational equations, rather than complicated interactions between baryons and dark matter.  There is a long history of connecting non-analytic dependence with light degrees of freedom, and one is quickly led to think about soft gravitational radiation as a possible source for MOND like effects.   We will argue that there is indeed such an effect, but two different methods of calculating it lead to the conclusion that it is both very small and has the wrong dependence on parameters to reproduce any form of MOND.

The basic idea underlying the effect is that gravitons of any arbitrarily low energy carry the same quantized units of helicity.  The rotation of stars around the galactic center\footnote{For simplicity, we consider only spiral galaxies and circular orbits.} leads to a bias in the polarization of the emitted soft gravitons.   We will argue that the emission of $N$ soft gravitons will lead to a loss of angular momentum of order $\sqrt{N}$ in $\hbar = 1$ units.  We will argue that for orbits of large radius the amount of angular momentum radiated ($\Delta L/L$) is much larger than the radiated energy $\Delta E/E$.  
To a distant observer, after the radiation has passed by, the motion of the stars will indicate a larger angular momentum than that which contributes to the local energy of the rotating system\footnote{We are working in a fixed coordinate system, the approximate rest frame of all the non-relativistic objects in the galaxy, so the notion of localized energy density in that coordinate system makes sense.}.  Since essentially no energy has been radiated, the equation for energy balance will be
\begin{equation} E = \frac{(L - \Delta L)^2}{2mr^2} - \frac{M(r) m}{r} . \end{equation}  Here and henceforth we use units where $\hbar = c = G_N = 1$.  $m$ is the mass of a particular star, in a circular orbit of radius $r$. $M(r)$ is the mass of the rest of the galaxy inside that radius.  The emitted angular momentum $\Delta L$ carries essentially zero energy.

In the MOND interpretation of the world, $M(r)$ represents only baryonic matter.  Our calculations apply not only to such a scenario but also to one in which the galaxy is filled with dark matter in the conventional $\Lambda-$CDM scenario.  They represent a real consequence of General Relativity, and if they gave a large correction to Newtonian expectations they would have to be taken into account.   Assuming $L \gg |\Delta L$ they could be mimicked by adding a fake dark matter density
\begin{equation} \Delta M = \omega \Delta L N_{star} , \end{equation} where $N_{star}$ is the number of stars in a small tube surrounding the orbit.

In the next section, we will calculate $\Delta L$ using linearized classical gravity.  After that we will present a calculation based on Weinberg's resummation formula for soft graviton emission in a general scattering process\cite{weinberg}.  Weinberg shows that to lowest order in $G_N$ his calculation reduces to the linearized classical calculation.  However, for the values of $m,M$ and $r$ relevant to our problem, we are far from the regime where such an expansion is valid.  Thus, our second calculation is probably a more accurate reflection of the exact answer.  We find that {\it both} methods of calculation lead to a very tiny result, with peculiar dependence on the parameters.  They are both proportional to $T_{gal}$, the lifetime of the galaxy.   In a universe with the cosmological constant $\Lambda = 0$, the effect would eventually become observable.  However, for the apparent value of $\Lambda $ in our universe, most galaxies will have merged with the cosmological horizon before this effect becomes measurable.  

\section{Linearized General Relativity}

The calculation of linearized gravitational energy emission from a non-relativistic, gravitationally bound two body system appears as a solved problem in Landau and Lifshitz' text on classical field theory. The calculation was redone by Peters and Mathews\cite{pm}, and Peters also calculated the rate of emission of angular momentum in his thesis.  According to these authors, the rate of energy and angular momentum emission averaged over one orbit are given by

\begin{equation} dE/dt = \frac{32}{5} \frac{M^2 (r) m^2 (M(r) + m)}{r^5} , \end{equation}

\begin{equation} dL/dt = \frac{32}{5} \frac{M^2 (r) m^2 (M(r) + m)^{1/2}}{r^{7/2}}  = (M(r) + m)^{-1/2} r^{3/2} dE/dt .\end{equation}

Note that at large $r$ the rate of emission of angular momentum is much larger than the energy emission rate.   The fraction of angular momentum emitted over the lifetime $T$ of the galaxy is of order 
\begin{equation} \frac{\Delta L}{L} = \frac{T M^{5/2} m^2}{r^{7/2} m r^2 \omega} , \end{equation} where $\omega = \sqrt{M/r^3}$  is the frequency of rotation.  This gives

\begin{equation} \frac{\Delta L}{L} = \frac{T M^{2} m}{r^4} . \end{equation}  The current age of the universe is $10^{61}$ in Planck units and galaxies formed before a tenth of this period passed so we take $T \sim 10^{60}$.   $m$, a typical stellar mass is $\sim 10^{35}$. The typical galactic mass contained in a sphere whose radius is of order that of the galaxy,  is about $10^{49}$ and the typical radius is $10^{55}$.  Thus an orbiting star will lose only about a fraction $10^{-33}$ of its angular momentum over its history so far.  In order to lose an order $1$ fraction we thus have to wait about $10^{33}$ times the current age of the universe.  However, according to current data, there is a cosmological horizon at about $10^{61}$ and all but our local group of galaxies will have disappeared into the horizon long before this loss of angular momentum becomes apparent.

\section{The Weinberg Approximation}

In 1965, Steven Weinberg calculated the rate of emission of soft gravitons in an arbitrary scattering process of elementary particles.   The calculation requires one to choose a scale $E_m$ which serves to differentiate gravitons that one considers soft from those one considers hard.  In our problem we will take $E_m$ to be the inverse size of a typical star, so that we are considering gravitons for which the star is indistinguishable from a point, and Weinberg's formulae apply.  Note that $\omega$, the orbital frequency is the typical graviton frequency emitted in linearized approximation, so cutting off the total energy emitted in soft gravitons by something smaller than $E_m$, we are still allowing many gravitons with orbital frequency to be emitted. Weinberg's formula for the rate of graviton emission with energy $\leq E \leq E_m$ is
\begin{equation} \Gamma (\leq E) = (E/E_m)^B f(B) \Gamma_0 , \end{equation} where $\Gamma_0$ is the rate for the process with all graviton radiative corrections, both real and virtual, omitted. In our problem the "process" under consideration will be the scattering of a star from angle $\theta$ to angle $\theta + \delta \theta$ in its orbit.  The constant $B$ is given by 
\begin{equation} B = \frac{1}{2\pi} \sum \eta_m \eta_n m_n m_m \frac{1 + \beta_{nm}^2}{\beta_{nm}(1 - \beta_{nm}^2)^{1/2}} {\rm ln}\ (\frac{1 + \beta_{nm}}{1 - \beta_{nm}}) . \end{equation}
The double sum is over all particles participating in the reaction. $\eta_n = {\pm 1}$ depending on whether $n$ is an incoming or outgoing particle, and $\beta_{nm}$ is the relative velocity.  In our problem there is one ingoing and one outgoing particle and $\beta_{nm} \ll 1$.  Thus $B = \frac{m^2}{2\pi} \gg 1$.  

In elementary particle reactions, $B \ll 1$, and Weinberg shows that an expansion in powers of $B$ reproduces the linearized gravity result analogous to the one computed in the last section.  However, in our problem $B \gg 1$.  It's clear that in this limit we only get a significant rate if $E = E_m$.  

Weinberg does not estimate the angular momentum emission, but it is easy to use his formulae to get a crude estimate.  The S matrix element for a radiative transition with emission of $N$ gravitons is
\begin{equation} S^{rad} = S \prod_{r=1}^N (\frac{8\pi}{2 (2\pi)^3 |{\bf q}_r|})^{ 1/2} \times \sum_n \eta_n \frac{[{\bf p}_n \cdot {\bf \epsilon}^* ({\bf q}_r, h_r/2)]^2}{p_n\cdot q_r} . \end{equation}
$h_r$ is the graviton helicity and ${\bf \epsilon}$ are the photon polarization vectors whose product gives the graviton polarization tensor.  Using the fact that the graviton momentum is the difference of incoming and outgoing particle (star) momenta, the leading contribution to polarized emission is proportional to the difference of these momenta which is $m\omega r$.  One can think of this in particle language as an ${\bf L \cdot S}$ coupling of the orbital motion to the graviton spin operator. 

We can calculate the expectation value of the helicity emitted with any single graviton 
by treating each factor in the product above as the amplitude for the final state to contain a graviton with momentum ${\bf q}$ and a helicity state characterized by the photon polarization vector ${\bf \epsilon}$.  The third component of the helicity operator is $\pm \hat{{\bf q}}$, where the polarization vectors for the two signs of helicity are complex conjugates of each other.  Thus, if $\hat{\bf q} = \begin{pmatrix} \sin (\theta ) \cos (\phi ) & \sin (\theta ) \sin (\phi ) &\cos (\theta ) \end{pmatrix}$, then the expectation value of the emitted helicity per graviton per unit time, in the rest frame of the galaxy, is proportional to 
\begin{equation} \langle H_3 \rangle = \int d\Omega \cos (\theta) \sum_{{\hat{\bf q}}, \epsilon} |S|^2 h(\epsilon) , \end{equation}  where $S$ is the one body S matrix element above and $h(\epsilon)$ is the graviton helicity for that polarization vector. This is the one body expectation value, and the full expectation value of helicity emission is the sum over gravitons of this result.  Note that the S matrix element is just proportional to $\hat{\bf r}\cdot {\bf \epsilon}^*  \hat{\bf p}\cdot {\bf \epsilon}^* , $ where $\hat{\bf r}$ is the instantaneous angular position of the star at which the graviton is emitted, and $\hat{\bf p}$ the instantaneous direction of its velocity.  In Weinberg's S-matrix formula, this factor comes from noting that (in the gauge where the time component of $\epsilon$ vanishes) ${\bf p_{out}} ={\bf  p_{in} }+ mr{\bf \omega} $, where ${\bf \omega}$ is the angular velocity of the orbit.  The polarization vectors for the two helicity states are complex conjugates of each other, so the expectation value vanishes.  This is easy to understand from considerations of time reversal invariance.  The background orbital motion violates T invariance, but the instantaneous emission amplitude treats this violation as a first order perturbation.  Thus, since the full theory is time reversal invariant, this expectation value is the expectation value of a T odd operator in an invariant state.  

The expectation value of $H_3^2$ is of course nonzero, which means that the full $N$ graviton helicity emission probability is like a random walk of $N$ steps and will be proportional to $\sqrt{N}$.  The probability contains a factor $|{\bf q}|^{-3}$ for each graviton, which leads to an infrared divergence exactly the same as that for the total rate of the process with $\leq E_m$ energy emitted in soft gravitons.  This means that the helicity emission is dominated by the softest gravitons allowed in the process.  In a cosmological context, causality implies that we should not consider gravitons with wavelength larger than the size of the apparent horizon, so the emission is dominated by gravitons of frequency $T_{gal}^{-1}$ where $T_{gal}$ is the lifetime of the galaxy at the time at which it is observed.

The total angular momentum per unit time will thus be dominated by the radiative transition rate for $E = E_m$ multiplied by the square root of the number of gravitons emitted.  This factor is $\sqrt{E_m T_{gal}} $. The function $f(B)$ is defined by
\begin{equation} f(B) = \frac{1}{\pi} \int_{-\infty}^{\infty} ds \frac{\sin s}{s} e^{B\int_0^1 \frac{dw}{w} (e^{iws} - 1)} . \end{equation} For large $B$ the integral over $s$ is dominated by the region around $s = 0$ because the real part of the coefficient of $B$ in the exponential is negative, but vanishes at $s = 0$.  It's then easy to see that $f \sim B^{-1/2} \sim 1/m$.
We conclude that the rate of emission of angular momentum is
\begin{equation} dL_{mem}/dt \sim  \sqrt{E_m T_{gal}} m^{-1} . \end{equation}   
This looks nothing like the classical result.  The classical formula describes a coherent state of gravitons of frequency approximately equal to the orbital frequency $\omega$ and includes the orbital angular momentum of those gravitons.  Our formula is dominated by the emission of extremely low frequency gravitons whose orbital angular momentum is negligible compared to their helicity.  Note that it's remarkable that such low frequency gravitons can effect the galaxy's rotation at all, since gravitons of wavelength longer than the size of the orbit look like pure gauge excitations.  This is the reason that we suspect that a proper interpretation of this effect will be as a new form of gravitational memory, representing a relative rotation of the Minkowski coordinates at infinity, relative to those near the galaxy.  Obviously, more work needs to be done to understand this properly.

The total angular momentum emitted as "memory" is of order 
\begin{equation} \Delta L_{mem} \sim \sqrt{E_m / m^2} \int_{T/10}^T dt t^{1/2} \sim \sqrt{T^3 E_m/ m^2} . \end{equation}
For a star of radius equal to that of the sun, $E_m \sim 10^{-44}$, and $m \sim 10^{38}$, both in Planck units.  $T \sim 10^{61}$ and
\begin{equation} \Delta L_{mem} \sim 10^{31} . \end{equation}  This should be multiplied by the number of stars in the galaxy. Recall that the classical angular momentum emission is 
\begin{equation} \Delta L =  \frac{32}{5} \frac{M^2 (r) m^2 (M(r) + m)^{1/2}}{r^{7/2}} T_{gal} \sim (M)^{5/2} m^2 r^{-7/2} T_{gal} \sim 10^{125} \times 10^{76} 10^{- 193} 10^{61} \sim 10^{69} , \end{equation} per star at radius $r$.  We've taken $r$ to be of order the size of a typical galaxy and $M(r)$ a typical galactic mass.    

Even when we average both results over all the stars in the galaxy, the classical emission, over the current lifetime of the universe, is many orders of magnitude larger than the emission in the form of helicity of soft gravitons. The classical gravitons are emitted in a coherent state and they carry non-negligible orbital angular momentum, so the calculations of soft graviton emission do not apply to them.  If we take the time to be very long, the soft graviton calculation eventually dominates the classical emission formulae but this is much longer than the time it will take most galaxies to disappear into our cosmological horizon.

\section{Conclusions}

We have explored two mechanisms for emission of angular momentum from a galaxy in the form of gravitational radiation of negligible energy.  Both sources are very small and could not become relevant to galactic dynamics before most galaxies merge with the cosmological horizon.  Classical emission leads to an effect which can be modeled by a theorist unaware of its existence, by an additional radius dependent "dark mass"
in the galaxy, but the form of this fake matter distribution does not have the dependence on the actual mass to explain the Tully-Fisher relation.  It is also, as we've reiterated, extremely small.   The very low energy quantum radiation, does not depend on the details of galactic dynamics at all.  It depends only on the mass and radius of typical stars, and the total number of stars in the galaxy.  Furthermore, it is stochastic, and not correlated with the galactic angular momentum at all.  It is equally likely to speed up a galaxies rotation as it is to slow it down.  In a universe with a much smaller cosmological constant than our own, but galaxies and stars with the same general properties, the quantum effect would dominate.

Finally we want to emphasize that although our considerations were carried out in a particular reference frame, where Newtonian gravity is a good approximation to General Relativity, our basic idea should have a covariant formulation.  As our title suggests, we speculate that what we have discovered is, {\it in Minkowski space}, a novel form of gravitational memory\cite{strometal}, in which asymptotic coordinate systems rotate with respect to those fixed locally on a galaxy, due to the emission of angular momentum carried by soft gravitons.

%\vfill\eject
\vskip.3in
\begin{center}
{\bf Acknowledgments }\\
The work of T.Banks is {\bf\it NOT} supported by the Department of Energy, the NSF, the Simons Foundation, the Templeton Foundation or FQXI. The research of WF is based upon work supported by the National Science Foundation under Grant Number PHY-1620610. \end{center}

\end{document}